\documentclass[aps,pra,showpacs,amsmath,amssymb,amsfonts,lengthcheck,twocolumn,longbibliography,superscriptaddress]{revtex4-2}

\usepackage{graphicx}
\usepackage{dcolumn}
\usepackage{bm}
\usepackage{braket}
\usepackage{enumitem}
\usepackage[utf8]{inputenc}
\usepackage{amsmath}
\usepackage{amssymb}
\usepackage{amsthm}
\usepackage{float}
\usepackage[colorlinks=true,citecolor=blue,linkcolor=blue,urlcolor=blue]{hyperref}

\newcommand{\e}{\mathrm{e}}

\newcommand{\bla}{bla\\bla\\bla\\bla\\bla}

\begin{document}

\title{Quantum speed limit for measurement probabilities}

\author{Agung Budiyono}
\email{agungbymlati@gmail.com}
\affiliation{Research Center for Quantum Physics, National Research and Innovation Agency, South Tangerang 15314, Republic of Indonesia}
\affiliation{Department of Physics, University of Maryland, Baltimore County, Baltimore, MD 21250, USA}
\affiliation{Quantum Science Institute, University of Maryland, Baltimore County, Baltimore, MD 21250, USA}



\author{Sebastian Deffner}
\affiliation{Department of Physics, University of Maryland, Baltimore County, Baltimore, MD 21250, USA} 
\affiliation{Quantum Science Institute, University of Maryland, Baltimore County, Baltimore, MD 21250, USA}
\affiliation{National Quantum Laboratory, College Park, MD 20740, USA}

\date{\today}

\begin{abstract} 
Any protocol to process quantum information has to conclude with a measurement, aimed at producing a specific set of probabilities of measurement outcomes. In this work, we investigate the time, energy and importantly the genuine quantum resources necessary for transforming a set of measurement probabilities generated by a positive-operator-valued measure (POVM), to a target set of measurement probabilities. To this end, we first show that the speed of measurement probabilities, defined as the average rate of the surprisal of measurement outcomes, is constrained by the genuine quantum fluctuations contained in the measurement probabilities. Interestingly, this quantum speed limit can act as a witness for bipartite quantum correlations by selecting an optimal local projective measurement. Furthermore, we obtain a minimum time to transform an initial measurement probabilities to a target measurement probabilities, and apply this result to analyzing the cost of generating a local athermality in terms of genuine quantum uncertainty.  
\end{abstract} 

\maketitle       

\section{Introduction}

The advent of quantum information theory has led to the development of various quantum-information processing protocols that are much faster and more efficient than possible in classical physics. Most of these protocols leverage the peculiar parallelism of the quantum dynamics. However, this comes with constraints, namely the Heisenberg uncertainty principle imposes quantum speed limits (QSLs), which restrict the maximum rate at which a quantum state can change \cite{Frey2016QINP,Deffner2017JPA}. Recent years have seen a flurry of research activity on QSL, see for instance Refs.~\cite{Mandelstam1945,Bhattacharyya1983JPA,Fleming1973,Anandan1990PRL,Pati1991PLA,Uhlmann1992PLA,Vaidman1992AJP,Uffink1993AJP,Margolus1998PD,Kupferman2008PRA,Levitin2009PRL,Jones2010PRA,Taddei2013PRL,Campo2013PRL,Deffner2013PRL,Pires2016PRX,Campaioli2018PRL,OConnor2021PRA,Poggi2021PRXQ,Hamazaki2022PRXQ,Deffner2022EPL,Shanahan:2018aa,Shiraishi:2018aa,Ito:2018aa,Mohan2022PRA,Pintos2022PRX,Carabba:2022aa,H_rnedal_2022,Shrimali2024PRA}, which is rooted in the numerous applications of QSLs in quantum information processing \cite{Lloyd2000Nature,Ashhab2012PRA,Deffner2020PRR,Canvea2009PRL,Mukherjee:2013aa,Campbell2017PRL,Aifer2022NJP,Giovannetti2011NPhot,Alipour:2014aa,Campbell2018QST,Deffner2010PRL,Mukhopadhyay2018PRE,Funo2019NJP,Deffner2025QST,Mohan2022NJP,Campaioli2022NJP,Allan2021Quantum,Pratapsi2025QST,Marvian2016PRA_2,Brown:2016aa,Tripathy:2025aa}. Yet, it is still not entirely settled to what extend the QSL is a unique feature of quantum dynamics \cite{Deffner2017NJP,Okuyama:2018aa}, which is why dedicated analyses have explored the relationship between speed and genuine quantum resources \cite{Mondal2016PLA,Marvian2016PRA,Budiyono:2025aa}. 

While QSLs originally define the minimum time needed for a quantum state to transition to a target state, standard approaches based on distances in state space do not always capture intuitive notions of speed of physical processes. For example, two states may be perfectly distinguishable geometrically, yet possess identical or near-identical expectation values for relevant physical observables. This limitation highlights the necessity for exploring different definitions of speed tailored to specific, realistic physical situations \cite{Mohan2022PRA,Pintos2022PRX,Carabba:2022aa,H_rnedal_2022,Shrimali2024PRA}. To this end, we developed a QSL for expectation values of an observable in terms of the asymmetry of the state relative to the observable in Ref.~\cite{Budiyono:2025aa}.

However, it is important to note that the quantum expectation value provides only limited information about the quantum statistics, since it represents only the first moment of the measurement probability. Fundamental information-theoretical quantities such as entropic measures or the Fisher information requires complete information about the measurement probabilities. As an important example, consider a situation in which we wish to determine the deviation from a thermal state with a sequence of measurements. Quantifying such ``athermality'' then necessitates the full knowledge of the measurement probabilities. The natural question arises how quickly measurement probabilities can evolve under quantum dynamics.  

In the present analysis, we first introduce a speed of measurement probabilities defined as the rate of the average of surprisal of measurement outcomes, and show that it is upper bounded by the genuine quantum uncertainty contained in the measurement probabilities. Hence, our QSL is genuinely quantum, in the sense that it vanishes for classical systems. The genuine quantum fluctuations are upper bounded by a nonadditive entropy measure, which for a rank-1 PVM (projection-valued measure) becomes equivalent to a measure of coherence. As an application, we then show that by selecting optimal local projective measurements, the QSL of measurement probabilities can serve as a witness for bipartite quantum correlations. Finally, we obtain a minimum time to transform an initial measurement probabilities to a target measurement probabilities and apply this framework to derive the minimum time necessary for generating local athermality in terms of genuine quantum uncertainty.  

\section{Speed from quantum uncertainty} 

Consider a quantum system with a finite-dimensional Hilbert space $\mathcal{H}$. Assume that the state $\varrho(t)$ evolves continuously in time, $t$. Performing a measurement described by a POVM $\mathcal{M}:\{M_k\}_{k=1}^d$, $M_k\ge 0$, $\sum_kM_k=\mathbb{I}$, over the state $\varrho(t)$, generates a set of measurement probabilities to get outcome $k$ at time $t$ following the Born's rule: $\mathcal{P}_\mathcal{M}^k(t)\equiv{\rm Pr}(k|\varrho(t),M_k)={\rm tr}\{M_k\,\varrho(t)\}$. In the following, our primary objective is to derive a relationship between the maximal speed of the measurement probabilities and the distinct quantum features encoded in the measurement probabilities.  

To this end, we need a suitable quantity which captures how quickly a general set of time-continuous normalized probabilities changes over time by quantifying its sensitivity to small time increments. Thus, we define
\begin{equation}
v_{\mathcal{M}}\equiv\frac{1}{2\,}\sum_k\left|\dot{\mathcal{P}}_\mathcal{M}^k(t)\right|\,,
\label{speed of measurement probabilities}
\end{equation}
where as always the dot denotes a derivative with respect to time. Note that in the definition of $v_\mathcal{M}$  we uniformly add the speed of all the elements of the set of the measurement probabilities \footnote{The factor $1/2$ is included to simplify later expressions.}. 

It is interesting to note that defining $v_\mathcal{M}$ as in Eq.~\eqref{speed of measurement probabilities} is a somewhat natural choice, since it is also identical to the average rate of change of the surpisal $I(\mathcal{P}_\mathcal{M}^k(t))\equiv -\log\mathcal{P}_\mathcal{M}^k(t)$. In fact, we simply have
\begin{equation}
v_{\mathcal{M}}=\frac{1}{2}\sum_k\left|\dot{I}(\mathcal{P}_\mathcal{M}^k(t))\right|\,\mathcal{P}_\mathcal{M}^k(t)\,.
\end{equation}
It is easy to see using Jensen's inequality that $v_{\mathcal{M}}\le \sqrt{F_t}/2$, where $F_t$ is the classical Fisher information about time imprinted in set of measurement probabilities \cite{Amari2000info.geometry}. Note that the classical Fisher information is always upper bounded by the quantum Fisher information $\mathcal{F}_t$ \cite{Paris2009IJQI} and thus also $v_{\mathcal{M}}\le \sqrt{\mathcal{F}_t}/2$.

For the following analysis, we now consider a scenario, in which we choose a particular POVM, and study the speed of measurement probabilities relative to this POVM measurement. Without loss of generality we will further restrict ourselves to unitary dynamics, and a generalization to arbitrary time-continuous quantum maps can be found in Appendix \ref{Extension of Proposition 1 to generic dynamics}. For unitary dynamics described by $U(t)$, Eq. \eqref{speed of measurement probabilities} can be expressed as 
\begin{eqnarray}
v_{\mathcal{M}}=\frac{1}{2}\,\sum_k\left|{\rm tr}\{H(t)[M_k,\varrho(t)]\}\right|,
\label{speed of measurement probabilities under unitary dynamics for general Schatten norm}
\end{eqnarray}
where $H(t):=-i\,\dot{U}(t)\,U(t)^{\dagger}$ is the Hamiltonian generating the unitary transformation $U(t)$. Now employing the H\"older inequality, $v_\mathcal{M}$ can be directly upper bounded as 
\begin{eqnarray}
v_{\mathcal{M}}\leq \|H(t)\|_p\,\sum_k\|[M_k,\varrho(t)]\|_q/2, 
\label{speed of measurement probabilities is upper bounded by energy and quantum fluctuation}
\end{eqnarray}
where $1/p+1/q=1$, $(p,q)\in [1,\infty)$ and $\|O\|_p:={\rm tr}\{|O|^p\}^{1/p}$, $|O|=\sqrt{OO^{\dagger}}$, is the Schatten $p$-norm.

As argued in Ref.~\cite{Aifer2022NJP}, the first term on the right-hand side of Eq.~\eqref{speed of measurement probabilities is upper bounded by energy and quantum fluctuation}, i.e. $\|H(t)\|_p$, can be interpreted as an energetic cost of implementing the unitary. It expresses that a higher speed, in general, requires a larger energy. On the other hand, the second term, i.e., 
\begin{eqnarray}
\mathcal{U}_\mathcal{M}^q\equiv \frac{1}{2}\,\sum_k\|[M_k,\varrho(t)]\|_q, 
\end{eqnarray}
quantifies the ``quantumness'' as a resource for the speed of measurement probabilities resulting from the noncommutativity between the state and the measurement. 

Hence, besides consuming the energy as is also the case for classical systems, the speed of measurement probabilities, $v_\mathcal{M}$ is also determined by a measure of distinct quantumness as a resource. Their product determines the truly required resource for the quantum speed of measurement probability. In particular, a large energy may nevertheless lead to a low speed of measurement probabilities when the quantumness resource is small. 
 
As a case of particular interest, we take $q=1$ (or $p=\infty$) to have 
\begin{eqnarray}
v_{\mathcal{M}}\leq\|H(t)\|_{\infty}\,\mathcal{U}_\mathcal{M}, 
\label{speed of measurement probabilities is upper bounded by energy and quantum fluctuation}
\end{eqnarray}
where $\mathcal{U}_\mathcal{M}=\mathcal{U}_\mathcal{M}^1=\frac{1}{2}\,\sum_k\|[M_k,\varrho(t)]\|_1$ to ease notation. Here, $\|H(t)\|_{\infty}$, which is the largest absolute eigenvalue of $H(t)$, can be seen as the worst energetic cost of implementing the unitary protocol. Moreover, $\mathcal{U}_\mathcal{M}$ captures a form of genuine quantum uncertainty encoded in the measurement probabilities as follows. 

\subsection{From uncertainty to non-extensive entropy}

We continue with a closer look at the quantum uncertainty $\mathcal{U}_\mathcal{M}$. In particular, we show in Appendix \ref{relation between trace-norm noncommutativity and KD quasiprobability}, that we have 
\begin{eqnarray}
\mathcal{U}_\mathcal{M}=\sum_k\sup_{\{\Pi_{\mu}\}}\sum_{\mu}\left|{\mathfrak Im}\left[{\rm Pr}_{\rm KD}(\mu,k|\varrho(t))\right]\right|,
\label{expression of the genuine quantum uncertainty in terms of the extremal nonreality of the KD quasiprobability}
\end{eqnarray}
where ${\rm Pr}_{\rm KD}(\mu,k|\varrho(t)):={\rm tr}\{\Pi_{\mu}M_k\varrho(t)\}$ is a generalized Kirkwood-Dirac (KD) quasiprobability associated with the state $\varrho(t)$ relative to the POVM $\{M_k\}$ and a rank-1 PVM $\{\Pi_{\mu}\}$ \cite{Kirkwood1933PR,Dirac1945RMP,Lostaglio:2023aa,Arvidsson-Shukur:2024aa,Gherardini:2024aa}; the supremum is taken over the set of all the (redundant) rank-1 PVMs of the Hilbert space $\mathcal{H}$. Hence, the instantaneous speed of measurement probabilities is upper bounded by the total sum of the imaginary part of the KD quasiprobability associated with $\varrho(t)$ over $\{M_k\}$ and a rank-1 PVM $\{\Pi_{\mu}\}$ and optimized over all possible choices of the latter. 

Such nonreal values of the KD quasiprobability have been regarded as a signature of quantumness manifesting noncommutativity underlying quantum coherence \cite{Budiyono2023PRA,Budiyono2024JPA_2}, asymmetry \cite{Budiyono2023JPA,Budiyono2023PRA_2}, quantum correlation including entanglement \cite{Budiyono:2023.general.quantum.correlation,Budiyono2025PRA}, and disturbances in weak measurement \cite{Aharonov1988PRL,Wiseman2002PRA,Jozsa2007PRA,Dressel2012PRA}. It also captures a notion of epistemic restriction underlying uncertainty relation \cite{Budiyono:2021aa,Budiyono:2019aa}, and indicates quantum contextuality through the protocol of the weak measurement with postselection \cite{Kunjwal2019PRA}. Moreover, the KD quasiprobability can be observed in experiment using various schemes \cite{Lundeen2005PLA,Aharonov1988PRL,Wiseman2002PRA,Jozsa2007PRA,Johansen2007PLA,Vallone2016PRL,Cohen2018PRA,Wagner2024QST,Chiribella2024PRR,Doucet2026KDQ}.  

In previous work \cite{Budiyono2024JPA}, it was shown that $\mathcal{U}_\mathcal{M}$ can be further upper bounded as 
\begin{equation}
\mathcal{U}_\mathcal{M}\le\sum_k\sqrt{\mathcal{P}_\mathcal{M}^k(t)\,(1-\mathcal{P}_\mathcal{M}^k(t))}\equiv\mathfrak{S}(\{\mathcal{P}_\mathcal{M}^k(t)\}). 
\label{KD-nonreality coherence is upper bounded by the A entropy} 
\end{equation}
Note that $\mathfrak{S}(\{\mathcal{P}_\mathcal{M}^k(t)\})$ is a nonadditive entropic measure over the set of measurement probabilities, $\{\mathcal{P}_\mathcal{M}^k(t)\}$. Namely $\mathfrak{S}_\mathcal{M}$ is a Schur concave function of the measurement probabilities, vanishing when there is an element of the measurement probabilities equals to unity, and attaining its maximum value when the measurement probabilities is uniform. Hence, unlike the energetic resource which can be unbounded, the quantumness resource for speed of measurement probabilities is bounded from above determined by the dimension of the Hilbert space. 

For pure states, $\varrho(t)=\ket{\psi}\bra{\psi}$, the inequality in Eq.~\eqref{KD-nonreality coherence is upper bounded by the A entropy} becomes an equality when the POVM is a rank-1 PVM \cite{Budiyono2024JPA}. With this in mind, it can be argued that the entropy $\mathfrak{S}_\mathcal{M}$ quantifies the total measurement uncertainty generated by the POVM measurement $\{M_k\}$ over the state $\varrho(t)$, while $\mathcal{U}_\mathcal{M}$ is the genuine quantum part of the measurement uncertainty \cite{Budiyono2024JPA}. In particular, when the POVM $\{M_k\}$ commutes with the state $\varrho(t)$, i.e., $[M_k,\varrho(t)]=0$ for all $k$, $\mathcal{U}_\mathcal{M}$ vanishes even though $\mathfrak{S}_\mathcal{M}$ may still be finite. This is for instance the case when the state $\varrho(t)$ is given by a classical statistical mixture of the elements of the PVM (projective-valued measure) $\{M_k\}$ so that there is a classical uncertainty that is left unaccounted by $\mathcal{U}_\mathcal{M}$. With this interpretation, Eq.~\eqref{speed of measurement probabilities is upper bounded by energy and quantum fluctuation} shows that the speed of measurement probabilities is constrained by, i.e., takes as a necessary resource, the genuine quantum uncertainty or fluctuation contained in the measurement probabilities. 

\subsection{Speed and quantum coherence}

We now continue with a specific case, namely if the POVM is given by a rank-1 orthogonal PVM $\{\Pi_k\}_{k=0}^{d-1}$ of a $d$-dimensional Hilbert space, $\mathcal{H}$. We note that such a rank-1 PVM determines an orthonormal basis $\{\ket{k}\}$ of $\mathcal{H}$ such that $\Pi_k=\ket{k}\bra{k}$. In this case, the genuine quantum uncertainty defined in Eq.~\eqref{expression of the genuine quantum uncertainty in terms of the extremal nonreality of the KD quasiprobability} becomes 
\begin{equation}
\mathcal{U}_\Pi=\sum_k\sup_{\{\Pi_{\mu}\}}\sum_{\mu}\left|{\mathfrak Im}\left[{\rm tr}\{\Pi_{\mu}\Pi_k\varrho(t)\}\right]\right|
\label{quantum speed of measurement probability for rank-1 PVM is upper bounded by the KD-nonreality coherence}
\end{equation}
As argued in Ref.~\cite{Budiyono2023PRA}, $\mathcal{U}_\Pi$ is a faithful quantifier of coherence, called the KD-nonreality coherence, in the state $\varrho(t)$ relative to the reference basis $\{\ket{k}\}$ \cite{Budiyono2023PRA}. Combining this with Eq.~\eqref{speed of measurement probabilities is upper bounded by energy and quantum fluctuation}, for a rank-1 PVM measurement, the speed limit of measurement probabilities is upper-bounded by the quantum coherence relative to the measurement basis. 

Next, one can see from Eq.~\eqref{speed of measurement probabilities is upper bounded by energy and quantum fluctuation} that $v_\Pi$ can be made infinite by choosing a unitary with unbounded Hamiltonian generator, $\|H(t)\|_{\infty}=\infty$. However, in realistic situations, one may have access only to a specific set of parameters, leading to a bounded Hamiltonian $H(t)$. Noting this, without loss of generality, we restrict ourselves to the set of unitaries for which $\|H(t)\|_{\infty}=1$, and define QSL of measurement probabilities as, $v_\Pi^\mathrm{QSL}\equiv\sup_{\|H(t)\|_{\infty}=1}\{v_\Pi\}$. Formally, this is equivalent to dividing both sides of Eq.~\eqref{speed of measurement probabilities under unitary dynamics for general Schatten norm} by $\|H(t)\|_{\infty}$, and optimizing over the set of all unitaries on the Hilbert space. Therefore, $v_\Pi^\mathrm{QSL}$ can also be understood as the optimal speed of measurement probabilities under a unitary protocol relative to the most energetically expensive implementation of the unitary, and optimized over all the unitaries. 

Finally, we also have $v_\Pi^\mathrm{QSL}\leq \mathcal{U}_\Pi$.  Interestingly, we show in Appendix~\ref{Proof of equality for a single qubit with PVM measurement}, that for a single qubit and rank-1 PVM measurement, this inequality is tight, and the right-hand side is equal to the $l_1$-norm coherence in the state $\varrho(t)$ relative to the measurement basis \cite{Baumgratz2014PRL}. In Appendix~\ref{Proof of equality for a single qubit with PVM measurement}, we also obtain for a single qubit a complementarity relation for the speeds of measurement probabilities associated with three mutually complementary rank-1 measurement bases.  

\subsection{Witnessing bipartite quantum correlation}

We conclude this section with an application of the QSL \eqref{speed of measurement probabilities} to detecting bipartite quantum correlations. To this end, we consider a bipartite system ${\rm AB}$ with a state $\varrho^{\rm AB}(t)$ on a finite-dimensional Hilbert space $\mathcal{H}^{\rm AB}=\mathcal{H}^{\rm A}\otimes\mathcal{H}^{\rm B}$, where $\mathcal{H}^{\rm A(B)}$ is the Hilbert space of the subsystem ${\rm A(B)}$. As before, we assume that the bipartite state evolves under a global unitary $U^{\rm AB}(t)$. 

We then make a measurement described by a PVM $\{\Pi^{\rm A}_k\otimes\mathbb{I}^{\rm B}\}$ on $\mathcal{H}^{\rm AB}$, where $\{\Pi^{\rm A}_k\}$ is a rank-1 PVM on $\mathcal{H}^{\rm A}$, describing a local projective measurement on the subsystem A. We want to study the speed of the set of measurement probabilities generated by the measurement described by the PVM $\{\Pi^{\rm A}_k\otimes\mathbb{I}^{\rm B}\}$ over the state $\varrho^{\rm AB}(t)$ which reads 
\begin{equation}
v_{\mathcal{M}}=\frac{1}{2}\,\sum_k\left|{\rm tr}\left\{H^{\rm AB}(t)[(\Pi^{\rm A}_k\otimes\mathbb{I}^{\rm B}),\varrho^{\rm AB}(t)]\right\}\right|\,.
\label{speed of local measurement probabilities}
\end{equation}
Correspondingly, the QSL becomes
\begin{eqnarray}
v_{\mathcal{M}}\leq\|H^{\rm AB}(t)\|_{\infty}\,\sum_k\|[\Pi^{\rm A}_k\otimes\mathbb{I}^{\rm B},\varrho^{\rm AB}(t)]\|_1/2. 
\label{speed of local measurement probabilities is upper bounded by energy and local genuine quantum uncertainty}
\end{eqnarray}

Now, we consider a unitary $U^{\rm AB}(t)$ such that the infimum of the left-hand side of Eq.~\eqref{speed of local measurement probabilities is upper bounded by energy and local genuine quantum uncertainty} over all rank-1 local PVMs $\{\Pi^{\rm A}_k\}$ of $\mathcal{H}^{\rm A}$ is not vanishing, i.e.,
\begin{eqnarray}
 \mathcal{Q}_{\Pi}(\varrho^{\rm AB}(t))\equiv\inf_{\{\Pi^{\rm A}_k\}}\mathcal{U}_{\Pi^{\rm A}}(\varrho^{\rm AB}(t))\geq 0,
\label{nonvanishing speed of local measurement probability indicates quantum correlation}
\end{eqnarray}
where $\mathcal{U}_{\Pi^{\rm A}}(\varrho^{\rm AB}(t))=\sum_k\|[\Pi^{\rm A}_k\otimes\mathbb{I}^{\rm B},\varrho^{\rm AB}(t)]\|_1/2$ is the bipartite generalization of the quantum uncertainty discussed above. We can also write, 
\begin{equation}
\mathcal{Q}_{\rm A}(\varrho^{\rm AB}(t))=\inf_{\{\Pi^{\rm A}_k\}}\sum_k\sup_{\{\Pi^{\rm AB}_l\}}\sum_l\left|{\mathfrak Im}\left[{\rm Pr}_{\rm KD}(k;l|\varrho^{\rm AB}(t)\right]\right|. 
\label{general quantum correlation}
\end{equation}
As before, ${\rm Pr}_{\rm KD}(k;l|\varrho^{\rm AB}):={\rm tr}\{\Pi^{\rm AB}_l(\Pi^{\rm A}_k\otimes\mathbb{I}^{\rm B})\varrho_{\rm AB}\}$ is the KD quasiprobability associated with the bipartite state $\varrho^{\rm AB}$ relative to the local projector $\Pi^{\rm A}_k\otimes\mathbb{I}^{\rm B}$ and global rank-1 projector $\Pi^{\rm AB}_k$ of $\mathcal{H}^{\rm AB}$. 

It is easy to see that $\mathcal{Q}_{\rm A}(\varrho^{\rm AB}(t))$ quantifies the minimum noncommutativity between the bipartite state $\varrho^{\rm AB}(t)$ and any local projective measurements $\{\Pi^{\rm A}_k\otimes\mathbb{I}^{\rm B}\}$. In previous work \cite{Budiyono:2023.general.quantum.correlation,Budiyono2025PRA} it was shown that $\mathcal{Q}_{\rm A}(\varrho^{\rm AB}(t))$ is a bonafide measure of discord-like general quantum correlation for the bipartite state $\varrho^{\rm AB}(t)$ satisfying a set of desirable requirements \cite{Adesso2016JPA}. Namely, i) $\mathcal{Q}_{\rm A}(\varrho^{\rm AB}(t))$ vanishes if and only if $\varrho^{\rm AB}(t)$ is a classically correlated state given by a hybrid classical-quantum state $\varrho^{\rm AB}=\sum_kp_k\Pi^{\rm A}_k\otimes\sigma^{\rm B}_k$ where $\{p_k\}$ is a set of normalized probabilities and $\{\sigma^{\rm B}_k\}$ is a set of  states on $\mathcal{H}^{\rm B}$; ii) $\mathcal{Q}_{\rm A}(\varrho^{\rm AB}(t))$ is invariant under any local unitary operation; iii) $\mathcal{Q}_{\rm A}(\varrho^{\rm AB}(t))$ is nonincreasing under any completely positive trace preserving map on the subsystem ${\rm B}$; and iv) ${Q}_{\rm A}(\varrho^{\rm AB}(t))$ reduces to the entanglement measure for pure state. 

In fact, for a bipartite pure state $\ket{\psi(t)}^{\rm AB}$, we also have a closed expression as
\begin{equation}
\mathcal{Q}_{\rm A}(\varrho^{\rm AB}(t))={\rm tr}_{\rm A}\{(\varrho^{\rm A}(t)-\varrho^{\rm A}(t)^2)^{1/2}\}=\mathfrak{S}(\varrho^A(t)), 
\end{equation}
where $\varrho^{\rm A}(t)={\rm tr}_{\rm B}\{\ket{\psi(t)}\bra{\psi(t)}^{\rm AB}\}$. As noted above, $\mathfrak{S}_\Pi$ is a Schur concave function of the reduced density operator, hence, it is an entanglement monotone for the purified state $\ket{\psi(t)}^{\rm AB}$ \cite{Budiyono2025PRA,Vidal:2000aa}. From Eq.~\eqref{nonvanishing speed of local measurement probability indicates quantum correlation}, it is then clear that when the minimum of the speed of  measurement probabilities over all local rank-1 PVM measurement is non-vanishing, the global state $\varrho^{\rm AB}(t)$ is quantumly correlated. 

\section{Quantum speed limit time}

Having discussed the maximal rate, with which measurement probabilities can change, we now proceed to discuss the minimum time to obtain a set a specific set of measurement probabilities. As before, $H[\gamma(t)]$, is the Hamiltonian generating the unitary transformation $U[\tau,0;\gamma(t)]=\mathcal{T}_>e^{-i\int_0^{\tau} H[\gamma(t)]{\rm d}t}$, where $\mathcal{T}_>$ is the time ordering operator. Moreover, we now assume the Hamiltonian to be parametrized by a protocol $\gamma(t)$, that varies smoothly in time. Then the minimum time necessary to transform an initial set of measurement probabilities $\{\mathcal{P}_\mathcal{M}^k(0)\}$ to a target measurement probabilities $\{\mathcal{P}_\mathcal{M}^k(\tau)\}$ along a trajectory for the control parameters $\gamma(t)$ is lower bounded by 
\begin{equation}
\tau\ge \frac{\sum_k\left|\mathcal{P}_\mathcal{M}^k(\tau)-\mathcal{P}_\mathcal{M}^k(0)\right|}{2 \sqrt{\braket{\|H[\gamma(t)]\|_{\infty}^2}_{\tau;\gamma(t)}\braket{\mathcal{U}_\mathcal{M}^2}_{\tau;\gamma(t)}}}, 
\label{minimum time to get certain variational distance in terms of genuine quantum fluctuation}
\end{equation}
where $\braket{f(\gamma(t))}_{\tau;\gamma(t)}=\int_{0;\gamma(t)}^{\tau}{\rm d}t\,f(\gamma(t))/\tau$ is the time average of the function $f(t)$ along the trajectory $\gamma(t)$. See Appendix \ref{Proof of Proposition 2} for a proof. 

We emphasize that the lower bound in Eq.~\eqref{minimum time to get certain variational distance in terms of genuine quantum fluctuation} is proportional to the variational distance between the initial and final measurement probabilities, and inversely proportional to the time-average of energetic cost of implementing the unitary along the trajectory $\gamma(t)$ as intuitively expected. Most importantly, it is also inversely proportional to time-average of the genuine quantum fluctuation $\braket{\mathcal{U}_\mathcal{M}}_{\tau;\gamma(t)}$. Namely, the minimum time needed to transform a state to another state with a given targeted variational distance between the measurement probabilities, is determined by the total genuine quantum fluctuation consumed along the trajectory. In particular, even for large values of $\braket{\|H(t)\|_{\infty}}_{\tau;\gamma(t)}$ which represents a large amount of energetic cost, the QSL time  can be very large when the time-average genuine quantum fluctuations consumed along the trajectory, i.e., $\braket{\mathcal{U}_\mathcal{M}^2}_{\tau;\gamma(t)}$, is sufficiently small.  

\subsection{Optimal control protocols}

We further note that the choice of the POVM in Eq.~\eqref{minimum time to get certain variational distance in terms of genuine quantum fluctuation} is arbitrary. In particular, one can choose a POVM measurement which solves the variational expression for the trace distance between the initial and final states: $\sup_{\{M_k\}}\sum_k|\mathcal{P}_\mathcal{M}^k(\tau)-\mathcal{P}_\mathcal{M}^k(0)|=\|\varrho(\tau)-\varrho(0)\|_1$. In this case, the right-hand side of Eq.~\eqref{minimum time to get certain variational distance in terms of genuine quantum fluctuation} is just the minimum time necessary to prepare a state given an initial state unitarily. For this optimal measurement, Eq.~\eqref{minimum time to get certain variational distance in terms of genuine quantum fluctuation} restricts the minimum genuine quantum fluctuation accumulated along any trajectory parameterizing the unitary connecting two states in terms of the trace distance between the two states.  

The above results naturally leads to the question on the optimal unitary protocol or strategy, i.e., the optimal control protocol $\gamma(t)$ to attain a target set of measurement probabilities. In fact, given a unitary evolution with a Hamiltonian generator with a finite constant operator norm $\|H[\gamma(t)]\|_{\infty}$, the optimal strategy $\gamma(t)$ can simply be obtained by choosing $\gamma(t)$ that consumes the largest time-average genuine quantum fluctuation $\braket{\mathcal{U}_\mathcal{M}}_{\tau;\gamma(t)}$. 

Of particular interest is when we restrict the unitary so that $\|H[\gamma(t)]\|_{\infty}=1$ along the trajectory $\gamma(t)$, and choose a rank-1 PVM $\{\Pi_k\}$ measurement so that the genuine quantum uncertainty is equal to the quantum coherence. In this case, from Eqs.~\eqref{minimum time to get certain variational distance in terms of genuine quantum fluctuation} and \eqref{quantum speed of measurement probability for rank-1 PVM is upper bounded by the KD-nonreality coherence}, we obtain a simple relation for the minimum time to attain a target measurement probabilities in terms of optimal time-average quantum coherence accumulated along the trajectory as 
\begin{eqnarray}
\tau&\ge&\frac{\sum_k|\mathcal{P}_\Pi^k(\tau)-\mathcal{P}_\Pi^k(0)|}{2\,\sup_{\gamma(t)}\left\{\sqrt{\braket{\mathcal{U}_\Pi^2}_{\tau;\gamma(t)}}\right\}}. 
\label{minimum time to get certain variational distance in terms of quantum coherence}
\end{eqnarray}

\subsection{Athermality generation}

As an application, we discuss the minimum genuine quantum fluctuation necessary to prepare a nonequilibrium or athermal state with unitary maps starting from a locally thermal or Gibbs state. Athermality, that is the deviation from a thermal state, is a valuable resource for many thermodynamics tasks, see for instance Refs.~\cite{Rossnagel2014PRL,Campaioli:2023aa}. Recently, athermality has also been discussed as a resource for entanglement generation \cite{Oliveira-Junior:2024aa}, quantum phase sensing \cite{Yadin:2025aa}, and traded with nonstabiliserness \cite{Oliveira-Junior:2025aa}. 

Consider a thermal state $\varrho_{\rm eq}=e^{-\beta H}/Z$, with a reference thermal Hamiltonian $H=\sum_k E_k\,\Pi_k$, an inverse temperature $\beta$ and the partition function $Z=\sum_ke^{-\beta E_k}$. Then, one way to define the amount of athermality in a state $\varrho$ relative to the reference thermal state $\varrho_{\rm eq}$ is through the variational distance between the probabilities generated by the measurement described by the eigenbasis $\{\Pi_k\}$ of the reference Hamiltonian over the two states, i.e., $\{\mathcal{P}_{\Pi}^k={\rm tr}\{\varrho\Pi_k\}\}$ and $\{\mathcal{P}_{\rm eq}^k\equiv\e^{-\beta E_k}/Z\}$,
\begin{eqnarray}
\mathcal{A}(\varrho,\varrho_{\rm eq};\{\Pi_k\})=\frac{1}{2}\,\sum_k\left|\mathcal{P}_{\Pi}^k-\mathcal{P}_{\rm eq}^k\right|\,.
\label{variational athermality}
\end{eqnarray}

Note that starting from a thermal state, we cannot prepare an athermal state using a unitary evolution. Thus we need an ancillary interacting with the system of interest, and generate a local athermality in the system through a global unitary evolution of the total system-ancillary. For this purpose, we assume that the system $\rm S$ interacts with an ancilla $\rm E$, having a total state $\varrho^{\rm SE}(t)$ on a finite dimensional Hilbert space $\mathcal{H}^{\rm SE}$. The state of the system reads $\varrho^{\rm S}(t)={\rm tr}_{\rm E}\{\varrho^{\rm SE}(t)\}$. We choose for the POVM the following set of local projectors $\{M_k\}=\{\Pi^{\rm S}_k\otimes\mathbb{I}^{\rm E}\}$, where $\mathbb{I}^{\rm E}$ is the identity operator on $\mathcal{H}^{\rm E}$ and $\{\Pi^{\rm S}_k\}$ is a PVM on the Hilbert space $\mathcal{H}^{\rm S}$. Now at time $t=0$, we choose a locally thermal state relative to $\{\Pi^{\rm S}_k\}$, i.e., $\varrho^{\rm SE}(0)$ such that ${\rm tr}_E\{\varrho^{\rm SE}(0)\}=\varrho_{\rm eq}^{\rm S}=\sum_ke^{-\beta E_k}\Pi^{\rm S}_k/Z$. As an example, $\varrho^{\rm SE}(0)$ can be taken as a so-called \emph{thermofield double state} \cite{cottrell2019JHEP}. We thus have $\mathcal{P}_{\rm eq}^k={\rm tr}\{(\Pi^{\rm S}_k\otimes\mathbb{I}^{\rm E})\varrho^{\rm SE}(0)\}={\rm tr}_S\{\varrho_{\rm eq}^{\rm S}\Pi^{\rm S}_k\}$. We evolve the state $\varrho^{\rm SE}(0)$ under a global unitary $U^{\rm SE}(t)$ to generate a local athermality in the system. 

Finally, using Eqs. (\ref{variational athermality}) in (\ref{minimum time to get certain variational distance in terms of genuine quantum fluctuation}) the minimum time to produce a local athermality $\mathcal{A}(\varrho^{\rm S}(\tau),\varrho^{\rm S}_{\rm eq};\{\Pi^S_k\})$ then reads,
\begin{eqnarray}
\tau&\ge&\frac{\mathcal{A}(\varrho^{\rm S}(\tau),\varrho^{\rm S}_{\rm eq};\{\Pi^S_k\})}{\sqrt{\braket{\|H^{\rm SE}[\gamma(t)]\|_{\infty}^2}_{\tau;\gamma(t)}\braket{\mathfrak{S}(\{\mathcal{P}_{\Pi^{\rm S}}^k(t)\})^2}_{\tau;\gamma(t)}}}.  
\label{minimum time to generate an athermality}
\end{eqnarray}
Here, $\mathcal{P}_{\Pi^{\rm S}}^k(t)={\rm tr}\{(\Pi^{\rm S}\otimes\mathbb{I}^{\rm E})\varrho^{\rm SE}(t)\}$ and we have used Eq.~\eqref{KD-nonreality coherence is upper bounded by the A entropy} to get $\mathcal{U}_{\Pi^S}(\varrho^{SE}(t))=\|[\Pi^{\rm S}_k\otimes\mathbb{I}^E,\varrho^{\rm SE}(t)]\|_1\le \mathfrak{S}(\{\mathcal{P}_{\rm \Pi^{\rm S}}^k(t)\})$. We conclude from Eq.~\eqref{minimum time to generate an athermality} that local measurement uncertainty quantified by the time-average local entropy $\braket{\mathfrak{S}(\{\mathcal{P}_{\rm \Pi^{\rm S}}^k(t)\})}_{\tau;\gamma(t)}$ is a resource for the generation of the local athermality. Reversing the time, Eq.~\eqref{minimum time to generate an athermality} can also be seen as the minimum time necessary starting from any state to arrive at a thermal state relative to a specific Hamiltonian. 

\section{Conclusions}   

In this work, we have discussed the speed of measurement probabilities defined as the average rate of surprisal of outcomes of a measurement. We have shown that the minimum time to attain target measurement probabilities under unitary state evolution is determined by the genuine quantum fluctuation contained in the measurement probabilities, which reduces to a coherence measure when the measurement is given by a rank-1 PVM. We have demonstrated the utility of this framework for witnessing bipartite quantum correlations and by identifying the cost for generating local athermality in terms of genuine quantum fluctuation. Our results clarify the interplay between the constraint imposed by the Heisenberg uncertainty principle on the speed of measurement probabilities and the genuine quantum fluctuation encoded in the measurement probabilities necessary for achieving quantum advantage. 

\acknowledgments{A.B. acknowledges the support from the Fulbright Visiting Scholar Program sponsored by the US Department of State during the completion the present work. AB would also like to thank ICTP Asean Network and PCS IBS for granting a short research visit. This work was partly supported by the {\it Lembaga Pengelolaan Dana Pendidikan} (LPDP) under the scheme of {\it Riset dan Inovasi untuk Indonesia Maju} (RIIM) managed by BRIN with the grant number: 61/II.7/HK/2024. S.D. acknowledges support from the John Templeton Foundation under Grant No. 63626.}

\appendix

\section{Extension of Eq. \eqref{speed of measurement probabilities is upper bounded by energy and quantum fluctuation} to generic time-continuous dynamics\label{Extension of Proposition 1 to generic dynamics}}

In this appendix, we discuss the generalization of the above speed of measurement probabilities to arbitrary time-continuous quantum dynamics. Let $\varrho^{\rm S}(t)$ denotes the state of the system with the Hilbert space $\mathcal{H}^{\rm S}$. Assume that the state of the system evolves following a generic time-continuous evolution: $\varrho^{\rm S}(t)=\Lambda_t(\varrho^{\rm S}(0))$, where $\Lambda_t$ is a superoperator. At time $t$, let $\ket{\psi(t)}^{\rm SE}$ be a purification of the system $\varrho^{\rm S}(t)$ in a larger Hilbert space $\mathcal{H}^{\rm SE}=\mathcal{H}^{\rm S}\otimes\mathcal{H}^{\rm E}$, where $\mathcal{H}^{\rm E}$ is the Hilbert space of the ancillary system for the purification, i.e., $\varrho^{\rm S}(t)={\rm tr}_E\{\ket{\psi(t)}\bra{\psi(t)}^{\rm SE}\}$, such that the purified system evolves unitarily $\ket{\psi(t)}^{\rm SE}=U^{\rm SE}(t)\ket{\psi(0)}^{\rm SE}$. Using the state purification, the speed of measurement probabilities can be written as 
\begin{equation}
\begin{split}
v_{\mathcal{M}}=&\frac{1}{2}\sum_k\left|\frac{{\rm d}}{{\rm d}t}{\rm tr}\{M^{\rm S}_k\varrho^{\rm S}(t)\}\right|\\
=&\frac{1}{2}\sum_k\left|\frac{{\rm d}}{{\rm d}t}{\rm tr}\{(M^{\rm S}_k\otimes\mathbb{I}^{\rm E})\ket{\psi(t)}\bra{\psi(t)}^{\rm SE}\}\right|\\
=&\frac{1}{2}\sum_k\left|{\rm tr}\{H^{\rm SE}(t)[(M^{\rm S}_k\otimes\mathbb{I}^{\rm E}),\ket{\psi(t)}\bra{\psi(t)}^{\rm SE}]\}\right|. 
\label{extension to generic dynamics step 1}
\end{split}
\end{equation}
Here, $H^{\rm SE}(t):=-i\dot{U}^{\rm SE}(t)\,U^{\rm SE}(t)^{\dagger}$ is the Hermitian generator of $U^{\rm SE}(t)$. 

Applying the H\"older inequality to Eq.~\eqref{extension to generic dynamics step 1}, we obtain
\begin{eqnarray}
v_{\mathcal{M}}\le\frac{1}{2}\|H^{\rm SE}\|_p\sum_k\|[(M^{\rm S}_k\otimes\mathbb{I}^{\rm E}),\ket{\psi(t)}\bra{\psi(t)}^{\rm SE}]\|_q,
\end{eqnarray}
where $1/p+1/q=1$, $(p,q)\in [1,\infty)$. Again, we consider the case when $q=1$ to have 
\begin{eqnarray}
v_{\mathcal{M}}\le\|H^{\rm SE}\|_{\infty}\sum_k\Delta_{M^{\rm S}_k}(\varrho^{\rm S}(t)),
\label{extension to generic dynamics step 2}
\end{eqnarray}
where $\Delta_{M^{\rm S}_k}(\varrho^{\rm S}(t))\equiv\sqrt{{\rm tr}\{(M^{\rm S}_k)^2\varrho^{\rm S}(t)\}-{\rm tr}\{M^{\rm S}_k\varrho^{\rm S}(t)\}^2}$ is the quantum standard deviation of $M^{\rm S}_k$ in $\varrho^{\rm S}(t)$, and in the last equality we have made use of a general relation which applies for any pure state and Hermitian operator: $\|[O,\ket{\psi}\bra{\psi}]\|_1/2=\Delta_O(\ket{\psi}\bra{\psi}):=\sqrt{{\rm tr}\{O^2\ket{\psi}\bra{\psi}\}-{\rm tr}\{O\ket{\psi}\bra{\psi}\}^2}$. For the case when $\{M^{\rm S}_k\}$ is a projective measurement so that $(M^{\rm S}_k)^2=M^{\rm S}_k$ for all $k$, then we have 
\begin{eqnarray}
\sum_k\Delta_{M^{\rm S}_k}(\varrho^{\rm S}(t))=\sum_k\sqrt{\mathcal{P}_\mathcal{M}^k(t) (1-\mathcal{P}_\mathcal{M}^k(\tau)}=\mathfrak{S}_M. 
\end{eqnarray}
As we discussed in the main text, $\mathfrak{S}_M$ is a nonadditive entropy capturing the uncertainty in the set of measurement probabilities $\{\mathcal{P}_\mathcal{M}^k(t)\}$. Equation~\eqref{extension to generic dynamics step 2} can thus be expressed as 
\begin{eqnarray}
v_{\mathcal{M}}\leq\|H^{\rm SE}\|_{\infty}\,\mathfrak{S}_M\,.
\end{eqnarray}

\section{The derivation of Eq. \eqref{expression of the genuine quantum uncertainty in terms of the extremal nonreality of the KD quasiprobability}\label{relation between trace-norm noncommutativity and KD quasiprobability}}

First, as shown Ref.~\cite{Budiyono2023PRA_2}, the Schatten $1$-norm of any normal operator $O$ on a finite-dimensional Hilbert space $\mathcal{H}$ can be expressed as the following variational problem,
\begin{eqnarray}
\|O\|_1:={\rm tr}\{\sqrt{OO^{\dagger}}\}=\sup_{\{\Pi_{\mu}\}}\sum_{\mu}|{\rm tr}\{\Pi_{\mu}O\}|,
\label{Lemma on the variational expression on the trace-norm asymmetry}
\end{eqnarray}
where the supremum is taken over the set of all rank-1 PVMs on the Hilbert space $\mathcal{H}$. Noting the fact that $[M_k,\varrho(t)]$ is a normal operator, its trace norm can therefore be expressed as, using Eq. \eqref{Lemma on the variational expression on the trace-norm asymmetry}, 
\begin{equation}
\begin{split}
\|[M_k,\varrho(t)]\|_1/2=&\sup_{\{\Pi_{\mu}\}}\sum_{\mu}|{\rm tr}\{\Pi_{\mu}[M_k,\varrho(t)]\}|/2\\
=&\sup_{\{\Pi_{\mu}\}}\sum_{\mu}|{\mathfrak Im}[{\rm tr}\{\Pi_{\mu}M_k\varrho(t)\}]|\\
=&\sup_{\{\Pi_{\mu}\}}\sum_{\mu}\big|{\mathfrak Im}\left[{\rm Pr}_{\rm KD}(\mu,k|\varrho(t))\right]\big|.
\end{split}
\end{equation}
\qed

\section{Example of a single qubit\label{Proof of equality for a single qubit with PVM measurement}}

Consider a system of a single qubit and assume that the POVM is given by a rank-1 PVM $\{\Pi_{k_0},\Pi_{k_1}\}$. We note first that such a rank-1 PVM determines to an orthonormal basis $\{\ket{k_0},\ket{k_1}\}$ on the two-dimensional Hilbert space $\mathcal{C}^2$. Let $\{\ket{k_0},\ket{k_1}\}$ comprises the north and south poles of the Bloch sphere, respectively. 

The Hermitian operator $H(t)$ generating the unitary $U(t)$ on $\mathcal{C}^2$ can in general be written as  
\begin{eqnarray}
H(t)=h_0\ket{h_0}\bra{h_0}+h_1\ket{h_1}\bra{h_1}.
\label{general Hermitian operator on 2-dimensional Hilbert space}
\end{eqnarray} 
Here, for simplicity, we have notationally omitted the time dependence on the right-hand side, $\{h\}=\{h_0,h_1\}$, are the eigenvalues, and the corresponding eigenvectors $\{\ket{h_0},\ket{h_1}\}$ can be expressed using the Bloch sphere parameterization as 
\begin{eqnarray}
\ket{h_0}&:=&\cos\frac{\alpha}{2}\ket{k_0}+e^{i\beta}\sin\frac{\alpha}{2}\ket{k_1};\nonumber\\
\ket{h_1}&:=&\sin\frac{\alpha}{2}\ket{k_0}-e^{i\beta}\cos\frac{\alpha}{2}\ket{k_1}, 
\label{complete set of basis in the x-y plane}
\end{eqnarray}
where $\alpha\in[0,\pi]$, $\beta\in[0,2\pi)$ are respectively the polar and azymutal angles of the Bloch sphere. We can assume, without loss of generality, that $|h_0|>|h_1|$. Let us restrict to the set of unitaries such that $\|H(t)\|_{\infty}=|h_0|=1$. Then, we obtain 
\begin{widetext}
\begin{eqnarray}
v_{\Pi}^{\rm QSL}&=&\sup_{\|H(t)\|_{\infty}=1}\sum_{k=k_0,k_1}\big|{\rm tr}\{H(t)[\Pi_k,\varrho(t)]\}\big|/2\nonumber\\
&=&\max_{\{(h_0,h_1)\in\mathbb{R}^2||h_1|\le|h_0|=1\}}\max_{\{(\alpha,\beta)\in[0,\pi]\times[0,2\pi)\}}|h_0-h_1||\braket{k_1|\varrho(t)|k_0}||\sin\alpha\sin(\beta+\phi)|\nonumber\\
&=&|\braket{k_0|\varrho(t)|k_1}|\max_{\{(h_0,h_1)\in\mathbb{R}^2||h_1|\le|h_0|=1\}}|h_0-h_1|=2|\braket{k_0|\varrho(t)|k_1}|=C_{l_1}(\varrho(t);\{\ket{k_0},\ket{k_1}\})\,.
\label{trace-norm asymmetry vs average noncommutativity for a single qubit}
\end{eqnarray} 
\end{widetext}
Here, $\phi=\arg\braket{k_0|\varrho(t)|k_1}$, the maximum over angular variables $(\alpha,\beta)\in[0,\pi]\times[0,2\pi)$ are attained for $\alpha=\pi/2$ and $\beta=\pi/2-\phi$, and the maximum over $(h_0,h_1)\in\mathbb{R}^2$, $|h_1|\le|h_0|=1$ is attained when $h_1=-h_0$. This means that the generator $H(t)$ of the unitary must have eigenvalues $\{-1,1\}$ so that we have $|h_0-h_1|=2$. 

Now, let us consider a specific set of Pauli unitaries having the following form $U_{\rm P}(t,0;\vec{n}_U(\gamma(t)))=\mathcal{T}e^{-i\int_0^{\tau}\vec{n}_U(\gamma(t))\cdot\vec{\sigma}{\rm d}t}$, where $\vec{\sigma}=(\sigma_x,\sigma_y,\sigma_z)^T$ is the vector of Pauli operators and $\vec{n}_U(\gamma(t))$ is a unit vector in three dimensional space. We note that the eigenvalues of $\vec{n}_U(\gamma(t))\cdot\vec{\sigma}$ are $\{-1,1\}$ so that $\|H(t)\|_{\infty}=\|\vec{n}_U(\gamma(t))\cdot\vec{\sigma}\|_{\infty}=1$. Hence, restricting to this set of Pauli unitaries, from Eq. \eqref{trace-norm asymmetry vs average noncommutativity for a single qubit}, we always have 
\begin{eqnarray}
v_{\Pi}^{\rm QSL}=C_{l_1}(\varrho(t);\{\ket{k_0},\ket{k_1}\}). 
\label{quantum speed of measurement probability for rank-1 PVM in a single qubit is upper bounded by the l1-norm coherence}
\end{eqnarray}

We then obtain the following complementarity relation for the speed of the measurement probabilities for a single qubit. Consider three mutually unbiased bases (MUB) for a single qubit: $\mathbb{X}:=\{\ket{x_+},\ket{x_-}\}$, $\mathbb{Y}:=\{\ket{y_+},\ket{y_-}\}$ and $\mathbb{Z}:=\{\ket{z_+},\ket{z_-}\}$, where $\ket{x_{\pm}}=\frac{1}{\sqrt{2}}(\ket{z_+}\pm\ket{z_-})$ and $\ket{y_\pm}=\frac{1}{\sqrt{2}}(\ket{z_+}\pm i\ket{z_-})$. In this case, the $l_1$-norm coherence in the state $\varrho(t)$ relative to these MUB bases satisfy the following complementarity relation \cite{Cheng2015PRA}: $C_{l_1}(\varrho(t);\mathbb{X})^2+C_{l_1}(\varrho(t);\mathbb{Y})^2+C_{l_1}(\varrho(t);\mathbb{Z})^2=2|\vec{r}_{\rm S}|^2$. Here, we have used the Bloch sphere representation of the state as $\varrho(t)=\frac{\mathbb{I}}{2}+\frac{1}{2}\vec{r}_{\rm S}\cdot\vec{\sigma}$, where $\vec{r}_{\rm S}$ is a vector in three dimensional space with $|\vec{r}_{\rm S}|\le 1$. Combining this with Eq. \eqref{quantum speed of measurement probability for rank-1 PVM in a single qubit is upper bounded by the l1-norm coherence}, we thus obtain the following complementarity relation among the associated speed of measurement probabilities generated by the corresponding three MUB rank-1 PVMs $\Pi_{\mathbb X}=\{\Pi_{x_+},\Pi_{x_-}\}$, $\Pi_{\mathbb Y}=\{\Pi_{y_+},\Pi_{y_-}\}$ and $\Pi_{\mathbb Z}=\{\Pi_{z_+},\Pi_{z_-}\}$, $\Pi_{i_{\pm}}=\ket{i_{\pm}}\bra{i_{\pm}}$, $i=x,y,z$, as:
\begin{eqnarray}
\left[v_{\Pi_{\mathbb X}}^{\rm QSL}\right]^2+\left[v_{\Pi_{\mathbb Y}}^{\rm QSL}\right]^2+\left[v_{\Pi_{\mathbb Z}}^{\rm QSL}\right]^2=2|\vec{r}_{\rm S}|^2. 
\label{complementarity relation for speed limit of three MUB PVMs for a single qubit}
\end{eqnarray}   

\section{Proof of Eq. \eqref{minimum time to get certain variational distance in terms of genuine quantum fluctuation}\label{Proof of Proposition 2}}

From Eqs. \eqref{speed of local measurement probabilities} and \eqref{speed of local measurement probabilities is upper bounded by energy and local genuine quantum uncertainty}, we first have 
\begin{eqnarray}
\frac{1}{2}\sum_k\big|\dot{\mathcal{P}}_\mathcal{M}^k(t)\big| \le \mathcal{U}_\mathcal{M}\,\|H(t)\|_{\infty}. 
\label{Proof of Proposition 6 step 2}
\end{eqnarray}
Integrating over time along a trajectory $\gamma(t)$ parameterizing the unitary $U[\tau,0;\gamma(t)]$ and using the triangle inequality, we have
\begin{equation}
\begin{split}
\frac{1}{2}\sum_k|\mathcal{P}_\mathcal{M}^k(\tau)-\mathcal{P}_\mathcal{M}^k(0)|&=\frac{1}{2}\,\sum_k\Big|\int{\rm d}t\,\dot{\mathcal{P}}_\mathcal{M}^k(t)\Big|\\
\leq\frac{1}{2}\sum_k\int{\rm d}t\Big|\dot{\mathcal{P}}_\mathcal{M}^k(t)\Big|&\le\int_{0;\gamma(t)}^{\tau}{\rm d}t\,\mathcal{U}_\mathcal{M}\,\|H(t)\|_{\infty}. 
\label{Proof of Proposition 6 step 3}
\end{split}
\end{equation}
Applying the Cauchy-Schwartz inequality to the right-hand side, we obtain Eq. \eqref{minimum time to get certain variational distance in terms of genuine quantum fluctuation} 
\begin{eqnarray}
\tau&\ge&\frac{\sum_k|\mathcal{P}_\mathcal{M}^k(\tau)-\mathcal{P}_\mathcal{M}^k(0))|}{2\,\sqrt{\braket{\mathcal{U}_\mathcal{M}^2}_{\tau;\gamma(t)}}\sqrt{\braket{\|H(t)\|_{\infty}^2}_{\tau;\gamma(t)}}}. 
\label{Proof of Proposition 6 step 4}
\end{eqnarray}  
\qed

\bibliography{QSL-measurement.probabilities}

\end{document}